\documentclass{article}
\usepackage{frascatiphys}
\usepackage{psfrag}
\usepackage{epsfig}
\begin{document}
\title{
CHIRAL DYNAMICS and $B\to 3\pi$ DECAY}

\author{
Ulf-G. Mei{\ss}ner\\
{\em Forshungszentrum J\"ulich, Institut f\"ur Kernphysik (Theorie)}\\
{\em D-52425 J\"ulich, Germany} \\
 and       \\
{\em Karl-Franzens Universit\"at Graz, Institut f\"ur Theoretische Physik}\\ 
{\em A-8010  Graz, Austria}
}
\maketitle
\baselineskip=11.6pt
\begin{abstract}
I discuss our knowledge of the scalar sector of QCD and how it impacts the
determination of the CKM angle $\alpha$ from the isospin analysis of
$B\to \rho\pi$ decay.
\end{abstract}
\baselineskip=14pt
\section{Introduction and motivation}
CP violation has been established experimentally in the K-- and B--meson
systems. In the Standard Model, this can be explained in terms of one
single phase, which leads to complex entries in the CKM matrix. 
The unitarity of this matrix may be represented in terms of various triangles,
one of them to be measured at the B--factories. Any violation of unitarity
would be a signal of physiscs beyond the Standard Model. However, to achieve
the required accuracy to really test the relation $\alpha + \beta + \gamma = \pi$,
where  $\alpha, \beta$, and $\gamma$ are the three angles of the triangle,
one has to be able to precisely calculate or eliminate the final--state
interactions (FSI) of the mesons generated in the various B--decays 
(more precisely, it is the strong phase generated by the FSI 
associated with  diagrams of the ``wrong'' weak phase 
that pose especial difficulty). Here, we will be
concerned with the decay $B\to 3\pi$, because  the isospin analysis possible
in $B\to \rho\pi$ decay allows to extract $\sin (2\alpha)$~\cite{LNQS,SQ}.
Recent observations, however, have triggered the question about a possible
``hadronic pollution'' in the $\rho\pi$ phase space. In particular, the E791
collaboration has found that half of the rate of $D^- \to \pi^- \pi^+\pi^-$
decay goes via the  $D^- \to \pi^- \sigma(500) 
\to \pi^- \pi^+\pi^- $ doorway state \cite{e791ex}. This measurement was also
considered as further evidence for a light scalar--isoscalar meson, the
elusive $\sigma$. Furthermore, it was shown in
ref.\cite{deA} that the inclusion of this channel can improve the
theoretical description of the ratio
\begin{equation}
\label{rhoratio}
{\cal R} = {{\rm Br}(\bar{B}^0 \to \rho^\mp \pi^\pm) \over
{\rm Br}({B}^- \to \rho^0 \pi^-)} = 2.7 \pm 1.2~,  
\end{equation}
measured at CLEO and BABAR. Note that ${\cal R} \simeq 6$ at tree level 
in naive factorization. Since there is on--going debate about the nature
of the $\sigma$, we will address here the following questions:
\begin{itemize}
\item[$\ast$]~What do we know about the scalar sector of QCD?
\item[$\ast$]~What  is its impact  on $B\to \rho\pi$ decay?
\end{itemize}
\section{The scalar sector of QCD}
\begin{figure}[th]
 \vspace{7.0cm}
\includegraphics{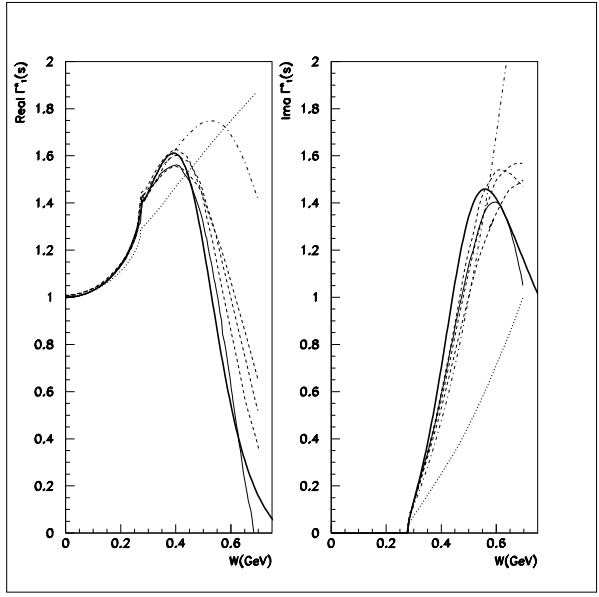}
\vspace{1.5cm}
 \caption{\it
      Real (left) and imaginary part (right) of the non-strange
      scalar form factor
      of the pion. Solid line: chiral unitary approach as discussed in
      the text, dotted/dash-dotted line: CHPT to one/two loops, dashed
      line: dispersive results.
    \label{sff} }
\end{figure}

\begin{figure}[t]
 \vspace{6.2cm}
\includegraphics{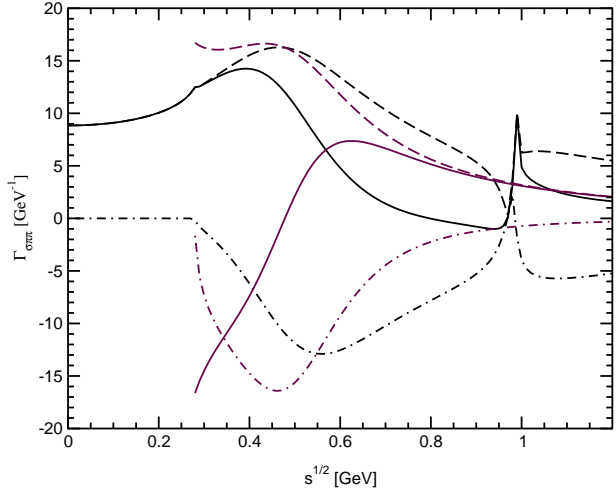}
\vspace{.65cm}
 \caption{\it
     The $\sigma\to\pi^+(p_+)\pi^-(p_-)$ form factor $\Gamma_{\sigma\pi\pi}$ as
     a function of $\sqrt{s}$, with $s=(p_+ + p_-)^2$.
     The real (solid line) and imaginary (dot-dashed line) parts of
     $\Gamma_{\sigma\pi\pi}$, as well as its modulus (dashed line), are
     shown. The curves which do not persist below physical threshold,
     $\sqrt{s}=2M_{\pi} \sim 0.27$ GeV, correspond to the form factor
     adopted in Ref.~\protect\cite{deA}, whereas the curves which extend to
     $s=0$ correspond to the form factor adopted here\cite{OM}. Both representations
     are normalized that their real parts agree at $\sqrt{s} = 0.478\,$GeV. 
    \label{sffbw} }
\end{figure}

The scalar--isoscalar sector of QCD is highly interesting because of its
vacuum quantum numbers, and its direct relation to the quark mass terms 
(explicit chiral symmetry breaking), the related $\sigma$--terms, and so on.
Its most distinct characteristics are the
very strong final--state interactions, signaled e.g. by the 
rapidly rising isospin zero, S-wave $\pi\pi$ phase shift $\delta_0^0 (s)$ or the
observation that the scalar pion radius, $\langle r^2_S\rangle_\pi
\simeq 0.6\,$fm$^2$, is sizeably bigger than the corresponding vector 
(charge) radius, $\langle r^2_V\rangle_\pi \simeq 0.4\,$fm$^2$.
Note also that there is no direct experimental probe with such
quantum numbers. Therefore, theoretical investigations using different tools
have been employed to deepen our understanding of this sector, these are
Chiral Perturbation Theory (CHPT), resummation schemes consistent with CHPT, unitarity, 
analyticity, $\ldots$ (like e.g. the chiral unitary approach\cite{ollrev})
and also dispersion relations. The following general results emerge: First,
a {\bf consistent} picture of the scalar  (pion and kaon) form factors is obtained,
and, second, all the light (non-strange and strange) scalar mesons 
are dynamically generated in a large class of resummation schemes 
(see e.g. ref.\cite{JOP}), although
this later topic is still vigorously debated\footnote{Consequently,
by ``$\sigma$'' we always mean a two--pion state with total isospin zero and
in a relative S--wave state, $(\pi \pi)_{{\rm S}}$, understanding its dynamical origin
in the strong pionic FSI for these quantum numbers.}.
For the impact on the $B\to \rho\pi$ decay, we fortunately only need the 
(non-strange) scalar form factor of the pion, $\Gamma_\pi (s)$ (or,
equivalently, the $\sigma \to \pi\pi$ vertex function $\Gamma_{\sigma\pi\pi}(s)$), 
defined via
\begin{equation}
\langle 0 | \hat{m} (\bar u u+\bar dd)| \pi^a (p) \pi^b(p')
\rangle = \delta^{ab} M_\pi^2 \, \Gamma_\pi (s)
 = {\cal N} \,  \Gamma_{\sigma\pi\pi} (s)~,~~s=(p´+p)^2~.
\end{equation}
The scalar form factor is shown in fig.~\ref{sff} for the various
theoretical approaches mentioned above. We will use here the result
of the chiral unitary approach \cite{OM}, which was  
successfully tested e.g. in $J/\Psi \to \phi \pi\pi (\bar KK)$ decays.
It is worth to point out that the scalar form factor constructed
in  \cite{OM} is systematically matched to the CHPT representation,
and it embodies by construction the coupled channel $\pi\pi / \bar K K$
dynamics. It is also consistent with the dispersive results of ref.\cite{DGL}.
Most importantly for the later discussion, we remark  that the form 
of the pion scalar form factor is very
different from a Breit-Wigner (BW) form with a running width, as used e.g. in
\cite{e791ex}, compare fig.~\ref{sffbw}. This apparent difference casts doubt
on the recent conclusions of refs.\cite{e791ex,deA}. The situation is completely 
different for the pion vector form factor entering the $\rho\pi$ intermediate
state - it can be described to good precision by a BW with running width.
More generally, the vector form factor can be reconstructed from unitarity and
analyticity and matched to CHPT. The resulting vertex function does not
differ significantly from a BW with running width (see ref.\cite{GM} for 
a detailed discussion on this point and corresponding figures).
\section{Extending the isospin analysis of $B\to \rho\pi$ decay}
Next, we wish to consider the impact of the $\sigma\pi$ channel on the
isospin analysis of $B\to \rho\pi$. Since in the initial state
the  B-meson has isospin  
$I_i=1/2$, and the final state $\rho\pi$ system
has $I_f = 0,1,2$, transitions with $|\Delta I|=1/2, 3/2, 5/2$
are possible. If one parameterizes the corresponding amplitudes 
$a_{bc} \equiv A(B^0 \to \rho^b \pi^c)$ (with $a,b = \{+,0,-\}$)
by $A_{|\Delta I|,I_f}$, one gets
\begin{eqnarray}
a_{+-} &=& 
{1\over 2\sqrt{3}} \left[A_{3/2,2}+A_{5/2,2}\right] + {1\over 2}
\left[A_{3/2,1}+A_{1/2,2}\right] + {1\over \sqrt{6}} A_{1/2,0}~, \nonumber \\
a_{-+} &=& 
{1\over 2\sqrt{3}} \left[A_{3/2,2}+A_{5/2,2}\right] - {1\over 2}
\left[A_{3/2,1}+A_{1/2,2}\right] + {1\over \sqrt{6}} A_{1/2,0}~, \nonumber \\
a_{00} &=& 
-{1\over \sqrt{3}} \left[A_{3/2,2}+A_{5/2,2}\right] 
+ {1\over \sqrt{6}} A_{1/2,0}~,
\end{eqnarray}
noting that
$A(B^0\to \pi^+\pi^-\pi^0)=f_+\, a_{+-} + f_-\, a_{-+} + f_0\, a_{00}$,
where $f_i$ is the form factor describing $\rho^i \to \pi \pi$.
Because of CKM unitarity, there are two independent weak phases,
a possible choice being
\begin{equation}
{V_{ub}^* \,V_{ud}^{}\over |V_{ub}^* \,V_{ud}^{}|} = \exp (i\gamma)~,~~
{V_{tb}^* \,V_{td}^{}\over |V_{tb}^* \,V_{td}^{}|} = \exp (-i\beta)~,~~
\alpha = \pi - \beta -\gamma~,
\end{equation}
leading to 
\begin{eqnarray}\label{physparam}
e^{i\beta}a_{+-} &=& T^{+-} \, e^{-i\alpha}+ P^{+-}~, \nonumber\\
e^{i\beta}a_{-+} &=& T^{-+} \, e^{-i\alpha}+ P^{-+}~, \nonumber\\
e^{i\beta}a_{00} &=& T^{00} \, e^{-i\alpha}+ P^{00}~,
\end{eqnarray}
from which $\sin(2\alpha)$ can be deduced, having made
the crucial assumption that the penguin is $|\Delta I|=1/2$, so that
\begin{equation}
\label{isopenguin}
P^{00} = {1\over 2}(P^{+-}+ P^{-+})~.
\end{equation}
With this penguin assumption, one has 10 parameters, 
which can all be determined from a Dalitz plot analysis.
As discussed in ref.\cite{GM}, 
there are three different sources of isospin violation (IV)
that could invalidate this analysis: 1) IV generates an additional amplitude of
$|\Delta I|=5/2$ character,  2) IV can distinguish the $f_i$, and
3) IV can generate penguins with $|\Delta I|=3/2$. Some of these effects can be 
mitigated in an empirically driven way, as long as $A_{5/2,2}$ and $A_{3/2,2}$
share the same weak phase. However, non-$|\Delta I|=1/2$ penguin effects,
be they electroweak penguin contributions or contributions consequent to
isospin-violating effects in the hadronic matrix elements of
$|\Delta I|=1/2$ operators, present an irreducible
hadronic ambiguity from the viewpoint of this analysis
(for a more detailed discussion, see ref.\cite{GM}).
Note that this is quite different to the isospin analysis 
of $B\to \pi\pi$ decay\cite{svg}.
Returning to the scalar sector, we remark that
$\sigma \pi$ contributes preferentially to the $\rho^0\pi^0$
final state and thus   can break the assumed penguin relation. However,
$B \to \sigma \pi$ has definite transformation properties under CP, so that
an extended isospin analysis is possible.
Defining $a^{\sigma}_{00}=A(B^0\to\sigma\pi^0)$, we have
\begin{equation}
\label{defsig0}
e^{i\beta} a^{\sigma}_{00} = T_{\sigma}^{00} e^{-i\alpha} + P_{\sigma}^{00}\;.
\end{equation}
$T_{\sigma}^{00}$ and $P_{\sigma}^{00}$ are unrelated to the parameters
of Eq.~(\ref{physparam}),
so that we gain four additional hadronic parameters. However, more
observables are present as well.
Including the scalar channel, we now have
$A_{3\pi}\equiv
A(B^0\to \pi^+\pi^-\pi^0)=f_+\, a_{+-} + f_-\, a_{-+} + f_0\, a_{00}
+ f_\sigma \, a^{\sigma}_{00}$, where $f_\sigma$ is the form factor
describing $\sigma\to\pi^+\pi^-$. For this extended analysis to be
useful, the $\sigma$ has to contribute significantly to $B^0,
\bar{B}^0$ decay. It is worth noting that any
discernable presence of the $B\to \sigma\pi$ channel in the
$B\to \rho\pi$ phase space falsifies the notion that
the ``nonresonant'' background can be
characterized by a single, constant
phase across the Dalitz plot~\cite{charles_nr}. 
\section{Evaluating $B\to \rho \pi$ in the presence of the $\sigma\pi$ channel}
\begin{figure}[t]
\hspace{3.0cm}
\psfrag{B}{{\large B}}
\psfrag{p}{{\large $\pi$}}
\psfrag{R = s, r}{{\large R = $\sigma \, , \, \rho$}}
\psfrag{Rpp}{{\large R$\pi\pi$}}
\psfrag{G}{{\large $\Gamma$}}
\epsfig{file=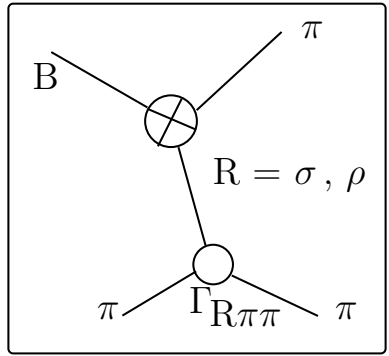,width=2in}
\vspace{0.1cm}
 \caption{\it
    Generic diagram for $B\to R\pi \to 3\pi$ decay, where $\Gamma_{R\pi\pi}$
    denotes the corresponding strong vertex function, $R=\sigma, \rho$. The crossed
    vertex symbolizes the weak $B\to R\pi$ transition. 
    \label{Bdecay}}
\end{figure}
Our starting point is the 
effective $|\Delta B|=1$ Hamiltonian for $b\to d q \bar{q}^\prime$
decay
\begin{equation}
{\mathcal H}_{\rm eff} = {G_F \over \sqrt{2}}\left[ \sum_{j=u,c} \, \lambda_j\,
\left( C_1 O_1^j + C_2 O_2^j \right) - \lambda_t \sum_{i=3}^{10} C_i O_i
\right]~,
\end{equation}
with $\lambda_q = V_{qb}^{}V_{qq}^{\star}$, $V_{ij}$ an element of the
CKM matrix and the operators are
ordered such that the Wilson coefficients obey
$C_1 \sim O(1), \, C_1 > C_2 \gg C_{3,\ldots , 10}$. We evaluate the
resonance contributions to $B\to 3\pi$ decay by using a product ansatz,
see fig.~\ref{Bdecay}.
For the requisite amplitudes, this means
\begin{equation}
A_R (B\to \pi^+ \pi^- \pi ) = \langle (R\to\pi^+\pi^-) \, \pi 
|{\mathcal H}_{\rm eff}| B \rangle 
=  \underbrace{ \langle \, R\pi\, |{\mathcal H}_{\rm eff}| B \rangle}_{\rm m.e.} 
\underbrace{\Gamma_{R\to\pi\pi}}_{\rm v.f.}~,
\end{equation}
where we compute the matrix element (m.e.) in factorization (including penguins)
in the same way it was done in refs.\cite{deA,deArho} for a crisp comparison (detailed
formulae can be found in ref.\cite{GM}). 
To ascertain the impact of the $B\to\sigma\pi$ channel to $B\to\rho\pi$ decay,
we combine the decay channels at the amplitude level,
${\mathcal M} = A_\sigma (B \to \pi^+ \pi^- \pi ) + A_\rho (B \to \pi^+ 
\pi^- \pi )\,$, and then integrate over the relevant three-body phase space
to   compute the effective  $B\to\rho\pi$ branching ratios.
The main new ingredient is the vertex function
(v.f.) for which we employ in the $\sigma\pi$ channel the scalar form factor 
discussed earlier and for the $\rho\pi$ mode the vector form factor from
ref.\cite{svghbo}. In table~\ref{tab:2body} we display the 
effective branching ratios for $B\to \rho\pi$ decay computed at
tree level (and also including penguins following ref.\cite{deArho}). 
\renewcommand{\arraystretch}{1.2}
\begin{table}[p]
\begin{center}
\caption{ \it 
Effective branching ratios (in units of $10^{-6}$) for
$B\to \rho\pi$ decay, computed at tree level.
The numbers in parentheses include
penguin contributions as well, after ref.\protect{\cite{deArho}}.
``BW'' denotes the use of the form factors of
refs.\protect{\cite{deA,deArho}},  whereas
``RW'' denotes the use of the vector form factor of
ref.\protect{\cite{babarbook}}.
Finally, ``$\ast$'' denotes the use of
the form factor  advocated here\protect{\cite{GM}}.
\label{tab:2body}
}
\vskip 0.1 in
\begin{tabular}{||l||c|c|c|c||}
    \hline
$\delta$ [MeV] (f.f.)
     &  $\bar B^0 \to \rho^- \pi^+$
     &  $\bar B^0 \to \rho^+ \pi^-$
     &  $\bar B^0 \to \rho^0 \pi^0$
     &  $B^-\to \rho^0\pi^-$   
\\ \hline \hline
200 (BW) & 15.1 (14.7) & 4.21 (4.24) & 0.508 (0.497) & 3.50 (3.68)
\\
  \hline
300 (BW)  &  16.4 (16.0) & 4.74 (4.76) & 0.918 (0.908) & 3.89 (4.10)
\\
   \hline
200 (RW)  & 15.1 (14.8) & 4.19 (4.21) & 0.468 (0.463)  &  3.49 (3.68)
\\
   \hline
300 (RW)  &  16.4 (16.0) & 4.69 (4.70) & 0.835 (0.831) &  3.87 (4.07)
\\
    \hline
200 ($\ast$) &  15.3 (14.9) & 4.26 (4.28) & 0.473 (0.467) &  3.49 (3.68)
\\
   \hline
300 ($\ast$) & 16.4 (16.0)  & 4.75 (4.76) & 0.865 (0.859) &  3.85 (4.06)
\\
   \hline
    \hline
$\delta$ [MeV] (f.f.)     &  $B^0 \to \rho^+ \pi^-$
     &  $B^0 \to \rho^- \pi^+$
     &  $B^0 \to \rho^0 \pi^0$
     &  $B^+\to \rho^0\pi^+$  \\
\hline
\hline
200 (BW)   &  15.1 (14.7) & 4.21 (4.15) & 0.508 (0.615)
         &  3.50 (3.68)
\\
  \hline
300 (BW)  &  16.4 (16.0) & 4.74 (4.67) & 0.918 (1.02)
         &  3.89 (4.10)
\\
   \hline
200 (RW)  &  15.1 (14.7) & 4.19 (4.13) & 0.468 (0.571)
         &  3.49 (3.68)
\\
   \hline
300 (RW)  &  16.4 (15.9) & 4.69 (4.62) & 0.835 (0.935)
         &  3.87 (4.07)
\\
   \hline
200 ($\ast$)  &  15.3 (14.8) & 4.26 (4.20) & 0.473 (0.576)
         &  3.49 (3.68)
\\
   \hline
300 ($\ast$)   & 16.4 (15.9)  & 4.75 (4.68) & 0.865 (0.963)
         &  3.85 (4.06)
\\
   \hline
   \hline
\end{tabular}
\end{center}
\begin{center}
\caption{Effective branching ratios (in units of $10^{-6}$) for
$B\to \sigma\pi$ and $B\to \rho\pi$
decay, computed at tree level.
The form factors are defined as in Table \protect{\ref{tab:2body}}.
                   \label{tab:cf}}
\vskip 0.1 in
\begin{tabular}{||l||c|c|c|c||c||}
    \hline
$\delta$  (f.f.)
& $B^-\to $ & $B^-\to$
& $\bar B^0 \to$ & $\bar B^0 \to$ &
${\cal R}$
\\ 
{}[MeV] & $\sigma\pi^-$ & $(\rho^0 + \sigma)\pi^-$ & $\sigma\pi^0$ & 
$(\rho^0 + \sigma)\pi^0$ & \\
\hline \hline
200 (BW) & 2.97 & 6.16 & 0.0258 & 0.516 &  3.1 \\
  \hline
300 (BW)  &  5.17 & 8.61 & 0.0457 & 0.940 & 2.5 \\
   \hline
200 (RW) & 2.97 & 6.19 & 0.0258 & 0.475 & 3.1 \\
  \hline
300 (RW)  &  5.17 & 8.62 & 0.0457 & 0.855 & 2.4 \\
   \hline
200 ($\ast$) & 4.11 & 7.61 & 0.0396 & 0.508 &  2.6 \\
  \hline
300 ($\ast$) &  7.01 & 10.7 & 0.0663 & 0.916 & 2.0 \\
   \hline
   \hline
\end{tabular}
\end{center}
\end{table}
We find that at tree level ${\cal R} \simeq 5.5$, and the inclusion
of penguin contributions lowers this value to ${\cal R}\simeq 5.1$.
Neither this ratio
nor the calculated branching ratios depend in any significant
way on the various vector form factors employed.
The $B \to \sigma\pi$ branching ratios are collected in table~\ref{tab:cf}.
The computed values of  ${\cal R} \simeq 2.0 \ldots 2.6$ are consistent
with the empirical value of ${\cal R}_{\exp} = 2.7 \pm 1.2$, albeit the
errors are large. The reduction in ${\cal R}$ is mostly due to the effect
of the $\sigma\pi$ channel on the $B^-$ decay mode.
Turning to $B\to \sigma\pi^0$ decay, we see that the contribution
of the $\sigma$ meson to $B^0 (\bar B^0) \to \rho\pi$ decay
is {\it much} smaller ---
with the scalar form factor we advocate, the effect is some
10\%. Interestingly the $\sigma$ has a tremendous impact on
$B^-\to\rho^0\pi^-$ decay (very similar to the
large effect in $D^- \to \pi^+ \pi^-\pi^-$), and a relatively modest one on
$\bar B^0\to\rho^0\pi^0$ decay.
Let us emphasize that we have realized our numerical analysis at tree level,
so that the precise numbers but not the trends will change when a more refined
analysis is performed. It is the relative size of the penguin contributions
in $\bar B^0\to\sigma\pi^0$
and $\bar B^0\to\rho^0\pi^0$ decay
which is of relevance to the isospin analysis to extract $\sin(2\alpha)$.
The presence of the $\sigma\pi^0$ final state in the $\rho^0\pi^0$
phase space can
break the assumed relationship, Eq.~(\ref{isopenguin}), between
the penguin contributions in $\rho\pi$ and thus
mimic the effect of isospin violation --- alternatively we can
expand the $\rho\pi$ analysis to include the $\sigma\pi$ channel.
It is worth noting that the 
$\sigma\pi^0$ and $\rho^0\pi^0$ contributions can, to some measure,
be distinguished. 
Certainly the $\sigma\pi^0$ and $\rho\pi^0$ contributions behave
differently under the cut on the invariant mass of the $\pi^+\pi^-$
pair. 
\begin{figure}[t]
\centering
\epsfig{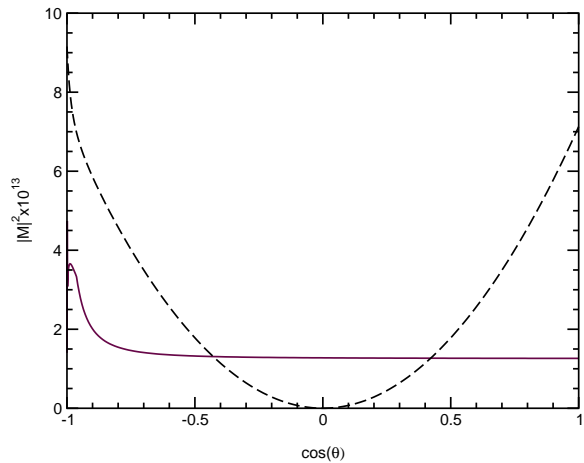}
\vspace{0.1cm}
 \caption{\it
             Absolute square of the matrix element,
             $|M|^2$, for $B^-\to\rho^0\pi^-$ decay (dashed line),
             and for $B^-\to\sigma\pi^-$ decay (solid line),
             as a function of $\cos\theta$ at $t=M_\rho^2$.
    \label{Bcut}}
\end{figure}
Moreover, making a cut on the helicity angle $\theta$,
ought also be helpful in separating the $\rho^0$ and $\sigma$
contributions.
This is illustrated in fig.~\ref{Bcut} for $B^- \to \rho^0/\sigma \, \pi^-$.
The $\rho^0\pi$ contributions roughly follow a $\cos^2(\theta)$ distribution, 
whereas the $\sigma\pi$ contributions are quite flat, save for
the bump resulting from the u--channel contribution 
$\sim \Gamma_{\sigma\pi\pi}(u)$.
Cutting on the helicity angle $\theta$ should also
help disentangle the contributions from some of the intermediate $B^*$\
and $B_0$ resonances. Such type of terms were claimed to be of importance
in ref.\cite{deArho}. The contributions of  these non--resonant intermediate
states to the $\rho\pi$ channels has recently been scrutinized
in ref.\cite{GT}, where it was shown that the energy dependence
of the intermediate heavy--meson propagator can lead to a drastic suppression
of such contributions and thus  the lowering of the value for ${\cal R}$
due to the $\sigma$ persists in such a refined analysis. 
We also point out that an additional contribution to the 
phenomenological value of ${\cal R}$, realized through
a diagram mediated by the $a_1^- (1260)$ meson, is proposed in
ref.\cite{paver} (it might turn out to be insignificant in a more
refined analysis for the same reasons just discussed for the $B^*$'s).

\section{Summary and outlook}
In this paper, we have scrutinized the role of the $\sigma$ meson
in $B\to\rho\pi\to 3\pi$ decay,
understanding its dynamical origin in the strong pion-pion final
state interactions in the scalar-isoscalar channel.
The presence of the $\sigma\pi^0$ contribution
in the $\rho^0\pi^0$ phase space is important in that it can
break the assumed relationship between the penguin amplitudes,
Eq.~(\ref{isopenguin}), consequent to an assumption of isospin
symmetry. In this, then, its presence mimics the effect of isospin
violation. The
salient results of our investigation can be summarized as follows:

\begin{itemize}

\item[i)] We have considered how SM isospin violation can impact
 the analysis to extract $\alpha$ in
  $B\to\rho\pi$ decay. Under the assumption that
  $|\Delta I| = 3/2$ and $|\Delta I| = 5/2$ amplitudes share the
  same weak phase, the presence of an additional amplitude of
  $|\Delta I|=5/2$ character, induced by isospin-violating effects,
  does not impact the $B\to\rho\pi$ analysis in any way. This is
  in contradistinction to the isospin analysis in $B\to\pi\pi$.
  Thus the isospin-violating effects of importance are those which
  can break the assumed relationship between the penguin contributions,
  Eq.~(\ref{isopenguin}).

\item[ii)]
  The scalar form factor
can be determined to good precision by
combining the constraints of chiral symmetry,
analyticity, and unitarity. The form factor we adopt describes
the appearance of the $f_0(980)$ as well, so that the shape of the
$f_0(980)$ contribution in $B\to f_0(980)\pi\to 3\pi$, e.g., should serve
 as a test of our approach. We emphasize that
the resulting
scalar form factor is very different from the commonly used Breit-Wigner
  form with a running width.  This is in stark contrast
  to the vector form factor, which is dominated by the $\rho$ resonance.
  In that case, one can construct  simple forms that fit the
  theoretical and empirical constraints.


\item[iii)] Remarkably, the impact of the $\sigma\pi$ channel on the
  ratio ${\cal R}$, cf. Eq.~(\ref{rhoratio}), is huge. The numbers
  we find for ${\cal R}$
  are in agreement with the empirical ones, given its sizeable
  experimental uncertainty. This underscores the suggestion made, as well
  as improves the calculations done, in Ref.~\cite{deA}. Our
  analysis is based on 
  {\em consistent} scalar and vector form factors. This conclusion
  persists if one includes non--resonant $B^*, B_0$ intermediate 
  states\cite{GT}.

\item[iv)] On the other hand, the impact of the $\sigma \pi$
  channel  on the $B\to\rho\pi$ isospin analysis is merely significant.
  Varying the cuts on the $\pi\pi$ invariant mass and helicity angle
  $\theta$ should be helpful in disentangling the various contributions.

\item[v)] We have shown that one can expand the isospin analysis
  to include the $\sigma\pi$ channel because it has definite
  properties under CP. This may be necessary if varying the
  cuts in the $\pi\pi$ invariant mass and helicity angle $\theta$
  are not sufficiently effective in suppressing the contribution
  from the $\sigma\pi^0$ channel in the $\rho^0\pi^0$ phase space.

\end{itemize}

\noindent
This work is merely a first step in
exploiting constraints from chiral symmetry, analyticity, and
unitarity in the description of hadronic B  decays.
In particular, the contribution
of the ``doubly''  OZI-violating strange scalar form factor
and its phenomenological role in factorization breaking
ought be investigated.
\section*{Acknowledgements}
I am grateful to Susan Gardner for a very pleasant collaboration on the
topics reported here and Jos\'e Antonio Oller and Jusak Tandean for useful
comments. I thank the organizers for their invitation and superbe organization.

\end{document}